\documentclass[10pt, conference, letterpaper]{IEEEtran}
\usepackage{amsmath}
\usepackage{tabularx,multirow}
\usepackage{graphicx,comment}
\usepackage[table]{xcolor}
\usepackage{url} 
\usepackage{psfrag,dsfont}
\usepackage{cite}
\usepackage{siunitx}
\usepackage{tabularx}
\usepackage{subcaption}
\usepackage{stfloats}

\usepackage[algoruled,medskip,dontprintsemicolon,linesnumbered,Algorithm]{algorithm}
\usepackage[noend]{algpseudocode}
\usepackage{mathtools}

\newcommand{\Bc}{\mathcal{B}}

\newcommand{\Cc}{\mathcal{C}}

\algnewcommand{\algorithmicand}{\textbf{ and }}
\algnewcommand{\algorithmicor}{\textbf{ or }}
\algnewcommand{\OR}{\algorithmicor}
\algnewcommand{\AND}{\algorithmicand}
\algnewcommand{\var}{\texttt}
\algrenewcommand{\algorithmicreturn}{\State \textbf{return}}
\usepackage{url}

\begin{document}

\title{Performance Analysis of C-V2I-based\\Automotive Collision Avoidance}

\author{\IEEEauthorblockN{M. Malinverno$^\ast$,
G. Avino$^\ast$, C. Casetti$^\ast$, C. F. Chiasserini$^\ast$,
F. Malandrino$^\ast$, S. Scarpina$^\dagger$}
\IEEEauthorblockA{$^\ast$Politecnico di Torino, Turin, Italy}
\IEEEauthorblockA{$^\dagger$TIM, Turin, Italy}
}
\maketitle

\begin{abstract}

One of the key applications envisioned for C-V2I (Cellular Vehicle-to-Infrastructure) networks 
pertains to safety on the road. Thanks to the exchange of Cooperative Awareness Messages (CAMs),
vehicles and other road users (e.g., pedestrians) can advertise their position, heading and speed 
and sophisticated algorithms can detect potentially dangerous situations leading to a crash. 
In this paper, we focus on the safety application for automotive
collision avoidance at intersections, and study 
 the effectiveness of its deployment in a C-V2I-based
 infrastructure. In our study, we also account for the location 
of the server running the application as a factor in the system design. Our simulation-based results, derived in 
real-world scenarios, provide indication on the reliability of algorithms for car-to-car and car-to-pedestrian
collision avoidance, both when a human driver is considered and when automated vehicles 
(with faster reaction times) populate the streets.
\end{abstract}


\section{Introduction}
In recent years, the development of vehicular network applications has
been attracting increasing interest from industries and researchers. A
critical field of application of vehicular networks is represented by
safety; indeed, in 2015 the number of people who lost their lives in
road traffic is more than 1.2 million \cite{GlobalStatus} and an
increasing trend in road casualties was observed in 2016 \cite{road-safety-2016}. A most
significant and, at the same time, challenging safety application is
\textit{collision detection}. One of the basic requirements for vehicles
running such an application is that they periodically send
\textit{Cooperative Awareness Message} (CAM) to a detector
\cite{sommer}. These messages are sent anonymously \cite{malandrino}
toward the \textit{base station} (BS) and contain information about
position, speed, acceleration and direction of the sender. The
\textit{collision detector} combines all the CAMs received by the
vehicles determining if any couple of vehicles is on a collision course.
If so, the drivers involved are immediately alerted. The communication
between vehicles and detectors happens through BSs that make
communication possible even in non-line-of-sight (NLoS) conditions,
e.g., due to buildings or other obstacles. 

The application can be extended also to vulnerable road users such as
pedestrians, whose smartphone can send CAMs to the detector. In this
way, both drivers and pedestrians are timely made aware of possible
life-threatening situations and can take proper action.

The purpose of this paper is to evaluate the performance of a system
for vehicle-with-vehicle and vehicle-with-pedestrian collision
detection when cellular vehicle-to-infrastructure (C-V2I)  is adopted
as a communication technology. In particular, we are mainly interested
in the number of collisions that could be avoided and in the
number of false positive alerts (i.e., alert messages referring to
situations of low or no danger, that the system delivers to the users). Indeed, a low number of false positive
alerts is essential in establishing user confidence in the reliability
of alerts received through the system.

The remainder of this paper is organized as follows: Section \ref{sec2}
reviews the research related to the automotive collision avoidance
application. Our reference scenario is introduced in
Section \ref{sec3}, while Section \ref{sec4} presents the design of
the automotive collision avoidance system, along with the 
detection algorithm. The description of the methodology for our
simulations and the output analysis technique are in Section \ref{sec5}.
Section \ref{sec6} contains the results obtained; the paper closes with
our conclusions and future research directions in Section \ref{sec7}.

\section{Related Work} \label{sec2}
There are several works in the literature that are related to safety
applications in the automotive domain (e.g., \cite{gallo}). Many of these works, such as
\cite{sightsafety} and \cite{wreckwatch}, propose collision avoidance
and collision detection applications that do not leverage any mobile
network infrastructure. In particular \cite{sightsafety} focuses on
collisions between vehicles and pedestrians in industrial plants. In this
case,  positioning is achieved using a combination of
GPS, MEMS and smart sensors, while the type of wireless communication to
the control center is not specified. In \cite{wreckwatch}, White et al.
propose a way to automatically detect a collision after it has occurred,
using smartphone accelerometers to reduce the time gap
between the actual collision and the first aid dispatch.

Our solution proposes a trajectory-based collision detection system
based on a state-of-the art algorithm that we enhanced to match our
needs.
The same base-algorithm has been used, in different flavors, in \cite{ptp} and \cite{wang}. \cite{ptp} offers a top-down and specification driven design of an adaptive, peer-to-peer based collision warning system, while \cite{wang} proposes a V2V-like approach. However, those two works offer little simulation results. In particular, \cite{ptp} only focuses on the collision avoidance algorithm, with little attention paid to implementation and network infrastructure. \cite{wang} provides some simulation results, but they tended to focus on the study of the algorithm parameters rather than applying the proposed solution to a realistic scenario. 

An attempt to  evaluate automotive forward collision
warning and avoidance algorithm (CW/CA) has been done in
\cite{evaluation}, where K. Lee et al. evaluated five different CW/CA
logics proposed by car makers. A very good survey of the strengths and
weaknesses of LTE as an enabler of vehicular communication technology is
\cite{survey}, where Araniti et al. also extended some of the
standardized safety messages used in this work.

\section{Reference Scenario} \label{sec3}

\begin{figure}[t!]
\centering
\includegraphics[scale=0.27]{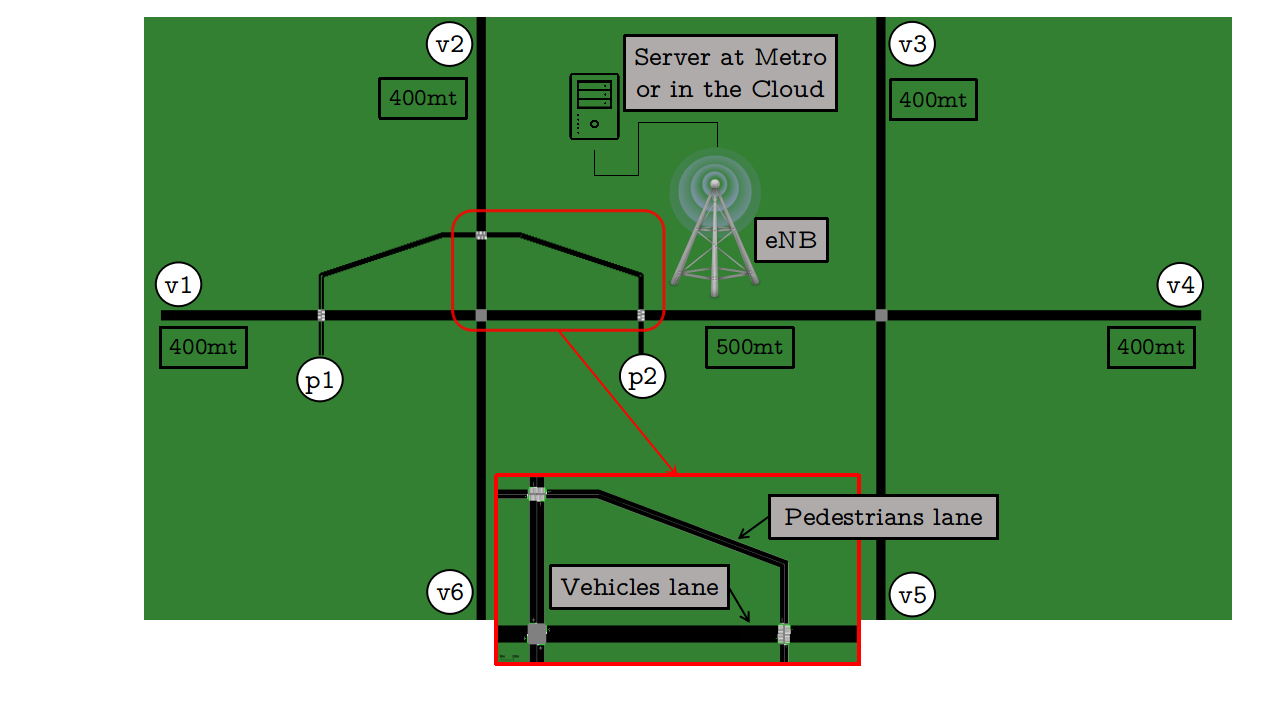}
\caption{Screenshot of the simulated scenario in SUMO.}\label{img:MapOverview}
\end{figure}

The reference topology we consider (depicted in Fig. \ref{img:MapOverview}) is an urban
area composed of three roads, crossing at two intersections, a
pedestrian lane and three pedestrian crossings. The intersections and
crossings are unregulated, which  makes collisions more
likely. The entities moving in the topology are vehicles and pedestrians. Each
of them is connected to the cellular infrastructure and uses the collision
avoidance service, i.e., we assume a penetration rate equal to 1. Vehicles are
equipped with  on-board units for C-V2I communications, whereas pedestrians carry a smartphone
with cellular connectivity. Both periodically send CAMs toward the
collision avoidance application server.

In particular, we consider an LTE network with 
an  eNodeB (eNB) located  at the center of the topology. The server
hosting the collision detector can be located at different points of the
network infrastructure, i.e., at the eNodeB itself or at more remote
network nodes. In order to study the difference in performance, we
consider two server deployments: at the Metro node (very close to the
eNB), in a multi-access edge computing (MEC) fashion, and in the Cloud (farther from the eNB). 
The choice of this urban topology 
allows us to have a simple but, at the same time, representative 
scenario, which closely mimics many real-world urban road layouts. 

In order to assess the performance of the collision detection service
in our scenario, we use the \textit{SimuLTE-Veins} simulator
\cite{simuLTEveins}, which leverages the mobility simulator SUMO
\cite{sumo}. 


\subsection{Populating the Scenario} \label{sub2.1}
We use a realistic mobility model and a realistic generation rate of
both vehicles and pedestrians. 
Vehicles have a maximum speed of $13.89$\,m/s (i.e., 50 km/h) and they follow a
straight path, i.e., there are neither left nor right turns at
junctions; pedestrians move with maximum speed of 2\,m$/$s on the
pedestrian lane, crossing the street at three different spots. Each
generated vehicle is randomly assigned to one of the six entry points
at the edge of the map (shown in Fig.~\ref{img:MapOverview} and marked
as \textit{v1}...\textit{v6}), while each vulnerable user is assigned to one of either ends of the pedestrian lane (\textit{p1} or \textit{p2}).
Following \cite{tmc-poisson}, vehicle arrivals are modeled as a Poisson process with
parameter $\lambda_v$; similarly, we model pedestrian arrivals with a
different Poisson process with rate $\lambda_p$. 

\begin{figure}[t]
\centering
\includegraphics[scale=0.4]{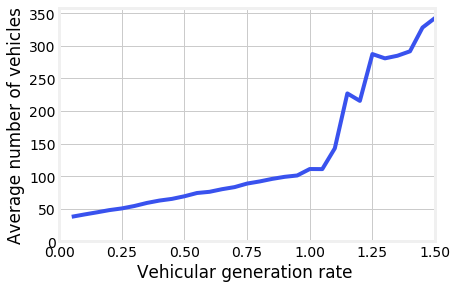}
\caption{Evolution of the average number of vehicles for $\lambda_p=0.2$ and $\lambda_v$ varying from $0$ to $1.5$.} \label{img:GenRate}
\end{figure}
\begin{figure*}
\centering
\includegraphics[scale=0.5]{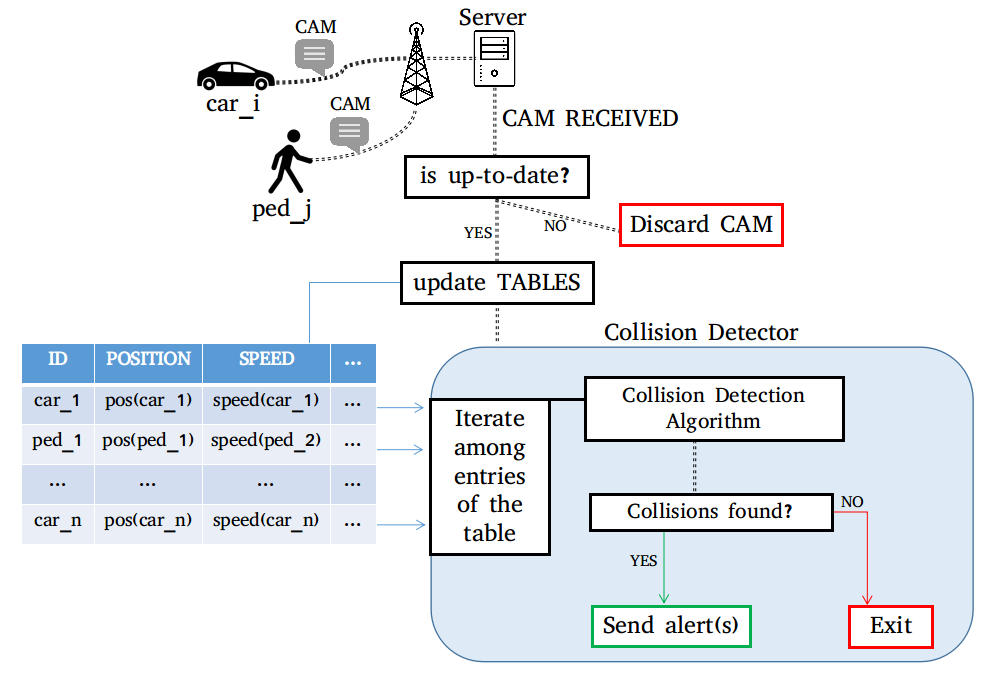}
\caption{Collision avoidance system.}\label{img:system}
\end{figure*}

In order to have reliable results and a realistic mobility pattern, we
apply the following methodology. We set both the vehicular generation
rate $\lambda_v$ and the pedestrian generation rate $\lambda_p$ in
such a way that the simulated scenario is stable; in other words, the
number of vehicles or pedestrians should never grow so high as to
yield the following situations: 
\begin{enumerate}
\item too many cars start clogging the intersection;
\item the long queues of low-speed vehicles result into a negligible
  number of collisions, and so the effectiveness of the automotive collision
  avoidance application cannot be correctly evaluated. 
\end{enumerate} 
The dependence of the number of vehicles and pedestrians from $\lambda_v$ and $\lambda_p$ is not linear because of the three crossings in which the two entities share the road occupancy.
To properly select the arrival rate of both cars and pedestrians, we
simulated the system when $\lambda_v$ varies from $0$ to $1.5$ whereas
$\lambda_p$ takes one of five possible values:
$0,0.05,0.10,0.15,0.20$. The results without pedestrians (i.e.,
$\lambda_p=0$) show that the value of~$\lambda_v$ for which the
average number of vehicles in the simulation grows linearly with the generation rate (i.e., it
is in the stability region) is between $0$ and $1.2$. The introduction
of pedestrians, however, has a significant impact: while with
$\lambda_p=0$ the maximum $\lambda_v$ allowing stability is $1.2$, by
increasing $\lambda_p$ to $0.2$, only values lower than $0.9$ ensure
stability. 
Fig.\,\ref{img:GenRate} shows how the average number of vehicles evolves for different values of $\lambda_v$ with $\lambda_p$ fixed to $0.2$.

Consequently, we set $\lambda_p$ to $0.1$ and $\lambda_v$ to
$0.7$. The value of $\lambda_v$ that we selected is consistent with
real-world measurements from the city of Turin, Italy \cite{5T-opendata}. 

\section{Design of the Collision Avoidance System} \label{sec4}
The collision detection system that we developed is shown in
Fig. \ref{img:system}. Below, we first introduce the collision
detection algorithm (Section \ref{sec4.1}), then we present the 
system design (Section \ref{sec4.2}). Finally, we show how to set the system
parameters (Section \ref{sec4.3}). 

\subsection{The Collision Detection Algorithm} \label{sec4.1} 
The core of the collision avoidance service is the detection
algorithm. Although our algorithm accounts for the vehicle
acceleration,   Algorithm
\ref{alg:collisions} presents a simplified version for the sake of clarity. 
Note that, being a generic trajectory-based algorithm, it can be
applied to  any kind of colliding entity (in our case both for vehicle-vehicle and vehicle-pedestrian collisions).
\begin{algorithm}[b!]
\fontsize{10pt}{14pt}\selectfont
\caption{Collision detection pseudocode\label{alg:collisions}
} 
\begin{algorithmic}[1]
\Require{$\vec{x_0},\vec{v},\Bc$} \label{line:input}
\State{$\Cc\gets\emptyset$} \label{line:init-c}
\State{$\vec{x}(t)\gets \vec{x_0}+\vec{v}t$} \label{line:pos}
\ForAll{$b\in\Bc$} \label{line:forb}
 \State{$\vec{x^b}(t)\gets \vec{x_0^b}+\vec{v^b}\cdot t$} \label{line:posb}
 \State{$\vec{d}(t)\gets \vec{x}(t)-\vec{x^b}(t)$} \label{line:dist}
 \State{$D(t)\coloneqq|\vec{d}(t)|^2\gets (\vec{v}-\vec{v^b})\cdot(\vec{v}-\vec{v^b})t^2+2(\vec{x_0}-\vec{x_0^b})\cdot\newline\textcolor{white}{iiiiiiiiiiiiii}\cdot(\vec{v}-\vec{v^b})t+(\vec{x_0}-\vec{x_0^b})\cdot(\vec{x_0}-\vec{x_0^b})$} \label{line:d2}
 \State{$t^{\star}\coloneqq t\colon\frac{\mathrm{d}}{\mathrm{d}t}D(t)=0\gets \frac{-(\vec{x_0}-\vec{x_0^b})\cdot(\vec{v}-\vec{v^b})}{|\vec{v}-\vec{v^b}|^2}$} \label{line:tstar}
 \If{$t^{\star}<0 \OR t^{\star}>t2c_t$} \label{line:check-past}
  \State{{\bf continue}}
 \EndIf
 \State{$d^{\star}\gets \sqrt{D(t^{\star})}$} \label{line:dstar}
 \If{$d^{\star}\leq s2c_t$} \label{line:check-dstar}
  \State{$\Cc\gets \Cc\cup \{b\}$} \label{line:add}
 \EndIf
\EndFor
\Return{$\Cc$} \label{line:return}
\end{algorithmic}
\end{algorithm}
The algorithm, which is based on \cite{collisiondetection}, is run when, after receiving a CAM, the detector determines that the sender of the message is on a collision course with another node. The collision detection algorithm requires as input (Line 0): 
\begin{itemize}
\item position and speed of the current vehicle, respectively
  identified by the two vectors~$\vec{x_0}$ and~$\vec{v}$; note that
  the speed vector also includes information on the heading;  
\item the latest CAM sent by each vehicle in the
  scenario stored in $\Bc$.   
\end{itemize}
In Line 1, the set~$\Cc$ of nodes with which the current entity could
collide is initialized and, in Line 2, the future position of the
current entity is evaluated for each future time instant. Then the
algorithm computes the position of each node $b\in\Bc$ that recently
sent a CAM  (Line 4) and the distance~$\vec{d}(t)$ between such a node
and the current entity (Line 5). In Line 6, we compute the square of
the distance~$D(t)\coloneqq|\vec{d}(t)|^2$; we do this to simplify the
subsequent computations. Since we are interested in the minimum value
of~$D(t)$, in Line 7 we compute~$t^{\star}$, defined as the time
instant at which the distance between the two entities is minimum
. If~$t^{\star}<0$, the two entities are getting farther apart, 
whereas, if~$t^{\star}$ is greater than a threshold~$t2c_t$ (where
$t2c$ stands for \textit{time to collision}), the minimum distance
will not be reached within $t2c_t$ from the current time. In both
cases, no action is required (Line 8). If~$t^{\star}$ is between 0
and~$t2c_t$, in Line 11 the minimum distance~$d^{\star}$ at which the
two entities will be at time~$t^{\star}$ is computed. The algorithm
compares $d^{\star}$ against a minimum threshold~$s2c_t$
(\textit{space to collision}): if~$d^{\star}$ is lower, then
vehicle~$b$ is added to set ~$\Cc$, otherwise the algorithm skips to
the next iteration of the cycle. 

Once all of the CAMs in set~$\Bc$ have been processed, the algorithm
returns the set~$\Cc$ of entities with which the current one is on a
collision course. If the set~$\Cc$ is empty, no action is taken, else
an alert message is sent to the current entity as well as  to all
entities in set~$\Cc$. 
We will discuss the setting of the thresholds $t2c_t$ and $s2c_t$ in Section~\ref{sec4.3}.

\subsection{System Description}\label{sec4.2} 
Looking at Fig. \ref{img:system}, we now specify how the other system
blocks work. 

The frequency at which CAMs are sent by each entity is 10\,Hz, which
is the maximum frequency allowed by the ETSI standard
\cite{etsi}. This high value allows the whole system to work with
updated information. Indeed, considering a lower frequency, e.g., 1\,Hz,
and a car moving at 13.89\,m$/$s (i.e., 50\,km$/$h), the error at the
server (ignoring the transmission delays) would be in the worst case
of~13.89\,m. Clearly, such a high error is not acceptable when dealing with a safety application.

The detector placed at the server is able to distinguish between CAMs
sent by pedestrians and CAM sent by vehicles. This gives us a double
advantage. First, when the server receives a CAM from a vehicle, it
looks for possible collisions with both cars and pedestrians, while on
the contrary, with a message sent from a pedestrian, the algorithm
skips the analysis for pedestrian-with-pedestrian collisions. The
second advantage involves the possibility to set different parameters
for the collision detection algorithm (i.e., $s2c_t$ and $t2c_t$),
according to the type of entity which sent the CAM. This allows a better performance of the algorithm, 
in terms of false positives and false negatives. 

Every time the detector receives a message, it checks if the message
is up-to-date: if so, the server stores the information of the CAM;
otherwise, the CAM is discarded. We set 0.8\,s as the threshold beyond which a
CAM is considered as stale and discarded. In the case of a fresh CAM,
the algorithm  checks if its sender is at risk of collision. To
improve the system efficiency, we also introduce a \textit{range of
  action}: only the entities within such a range  will be checked by
the algorithm as potential colliders. The radius varies according to the vehicle speed as follows:~$$\mbox{Radius}=\max\{\mbox{Speed}\cdot t2c_t, s2c_t\}$$

When the server detects a pair of entities on a course of collision,
they are warned by an alert message. In order to avoid an excessive
number of duplicated alerts, the collision detector does not generate the
same alert message more then one every second.

Thresholds and values used by the collision detector are summarized in Table \ref{tab:par} and discussed in the next section. 
\begin{table}[t!]
\centering
\caption{Collision detection parameters for vehicles and pedestrians.}
\label{tab:par}
\begin{tabular}{|c|c|c|cc}
\cline{1-3}
\multicolumn{1}{|c|}{\multirow{2}{*}{Parameter}} & \multicolumn{2}{c|}{Value} &  &  \\ \cline{2-3}
\multicolumn{1}{|c|}{} & Vehicle & Pedestrian &  &  \\ \cline{1-3}
$t2c_t$ & 10\,s & 5\,s &  &  \\ \cline{1-3}
$s2c_t$ & 5\,m & 2\,m &  &  \\ \cline{1-3}
Max CAM Age & 0.8\,s & 0.8\,s &  &  \\ \cline{1-3}
CAM frequency & 10\,Hz & 10\,Hz &  &  \\ \cline{1-3}
Alert max frequency & 1\,Hz & 1\,Hz &  &  \\ \cline{1-3}
\end{tabular}
\end{table}

\subsection{Sensitivity Study on Collision Thresholds} \label{sec4.3} 

Here we investigate the two parameters that mainly affect the performance of the collision detection system: $t2c_t$ and $s2c_t$.
Before delving into this study, we better detail their meaning below:
\begin{itemize}
\item $t2c_t$ is the \textit{time to collision threshold}. It
  introduces an upper bound on the \textit{time to collision} metric,
  i.e., the time gap needed for two entities to reach their mutual
  minimum distance. The higher this threshold, the more likely it is
  that a pair of entities are considered at risk of  collision. 
\item $s2c_t$: it is the \textit{space to collision threshold}. It is the upper bound to the distance at which two entities are at the \textit{time to collision}, to consider them at collision risk. Obviously, the higher the threshold, the more likely it is that a pair of entities is considered in collision course.  
\end{itemize}
Setting those two parameters has a big impact both on system accuracy
and on system efficiency: relaxing them means to let the algorithm
trigger too many alerts, even when candidates are not going to get
that close. Of course, in this case all the collisions are
correctly detected, but a large percentage of alerts refer to
low-danger situations. 
On the other hand, being too strict leads to the opposite situation in which the number of unnecessary alerts is drastically reduced, but a percentage of the collisions goes undetected or detected too late.
All the above are potentially dangerous for the driver. In particular,
for false positives it should be taken into account that an excessive
number of warnings may desensitize the driver, causing future alerts
to be ignored~\cite{ptp}. 

\begin{figure}[t!]
\subcaptionbox{Vehicle-with-vehicle: percentage of undetected or late-detected collisions\label{img:thresholds:a}}{\includegraphics[trim={0 0 3cm 0},clip,scale=0.23]{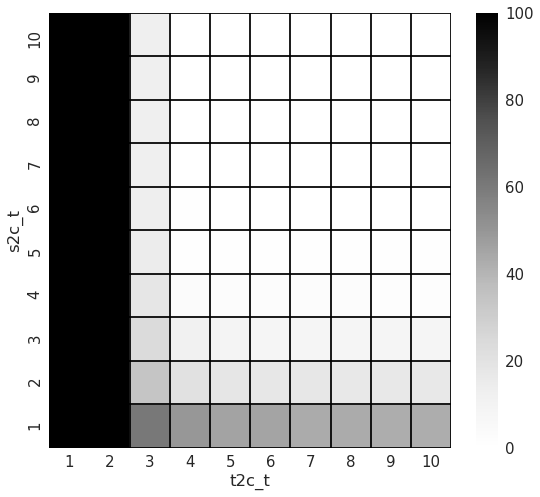}}
\subcaptionbox{Vehicle-with-vehicle: percentage of false positives\label{img:thresholds:b}}{\includegraphics[scale=0.23]{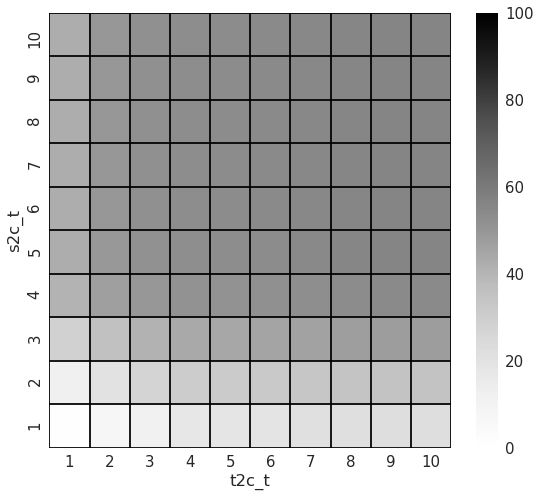}}
\captionsetup[subfigure]{position=b}
\subcaptionbox{Vehicle-with-pedestrian: percentage of undetected or late-detected collisions\label{img:thresholds:c}}{\includegraphics[trim={0 0 3cm 0},clip,scale=0.23]{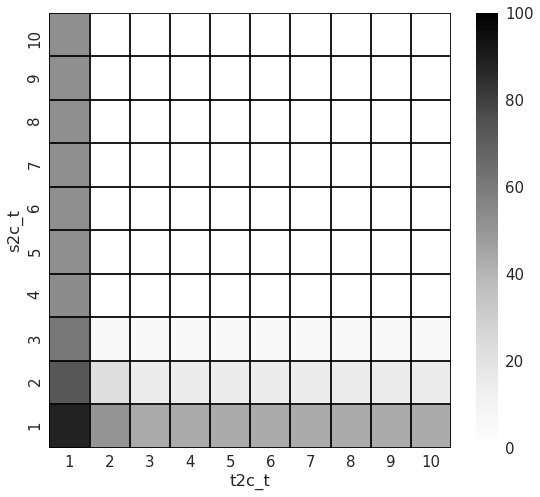}}
\subcaptionbox{Vehicle-with-pedestrian: percentage of false positives\label{img:thresholds:d}}{\includegraphics[scale=0.23]{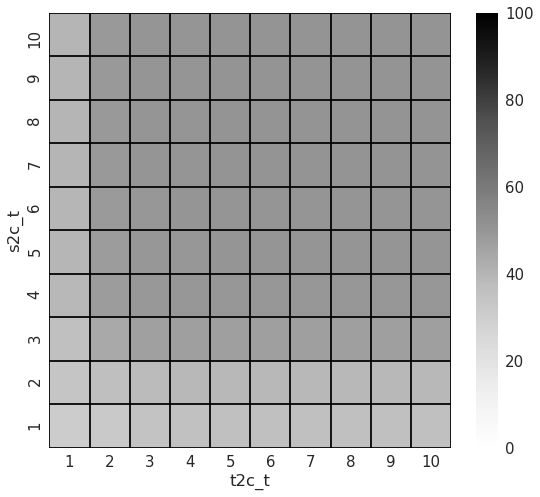}}
\captionsetup[subfigure]{position=b}
\centering
\caption{Thresholds analysis for vehicles traveling at 50~km/h.}
\label{img:thresholds}
\end{figure}

For these reasons, we undertook a study on the number of undetected or late-detected collisions as well as on false positives, as  functions of $s2c_t$ and $t2c_t$ (Fig. \ref{img:thresholds}).
Looking at the heatmap in Fig. \ref{img:thresholds:a}, it is clear
that for values of $t2c_t$ equal or lower than 3\,s, in a scenario
where vehicles travel at or around 50~km/h, the system is completely
unreliable. Indeed, considering the delays introduced by different
factors (e.g., processing time, human reaction, braking time), sending
the alert not earlier than 3\,s before the expected impact, does not
allow the driver to stop the vehicle and avoid the accident. Thus, the
$t2c_t$ value needs to be at least 4\,s. Looking at $s2c_t$ and
considering $t2c_t$ equal to 4\,s, any value greater than 3\,m ensures
the system to work efficiently, detecting in time the collisions
occurred. 

As mentioned before, the drawback of having high values of the
threshold is the number of false
positives. Fig. \ref{img:thresholds:b} shows that few false positives
can be obtained only with very low thresholds (both $t2c_t$ and
$s2c_t$ equal or lower than $2$), but this is completely unacceptable
given the high percentage of collisions not detected or detected too
late with such values. Thus, an interesting observation is that a high
number of false positives is the price to pay in order to realize a
reliable collision detection system. 

Looking at the case of vehicle-with-pedestrian alerts, we observe a
similar behavior. Fig. \ref{img:thresholds:c} shows that, even if the
trend is the same, a better performance is achieved by using smaller
values for the $t2c_t$ and $s2c_t$ thresholds. This happens because, in
the case of an alert, a pedestrian can stop almost instantaneously due
to her low speed. 
The above dynamic causes the total number of false positives to be in general higher than in the vehicle-with-vehicle case (Fig. \ref{img:thresholds:d}).

\section{Simulation Methodology} \label{sec5}

In this section, we first describe the approach we adopted to process
the simulation logs in order to derive the main metrics of interest. We
then detail how we discern whether alerts are received on time by the
involved entities so that the collision can be avoided.

\subsection{Processing the Simulation Logs}


We start by collecting the following information:
\begin{enumerate}
\item  the dynamics of vehicles and pedestrians
  (e.g., their position, speed and heading), using the SUMO \textit{Floating Car Data} output;
\item the vehicle-with-vehicle and vehicle-with-pedestrian collisions that occurred, through the SUMO error-log file;
\item all the alerts sent by the collision avoidance application,
  using the SimuLTE-Veins simulator.
\end{enumerate}

Then, through post-processing, we analyze, for each collision, when it
occurred and \textit{if} the corresponding alert message was generated. Furthermore, if the alert was correctly transmitted, we
also look at \textit{when} it was received and processed by 
the involved entities. 
In this way, we can determine if the vehicle had sufficient
time to brake before the impact.

\begin{figure*}[t!]
\centering
\includegraphics[width=1\textwidth]{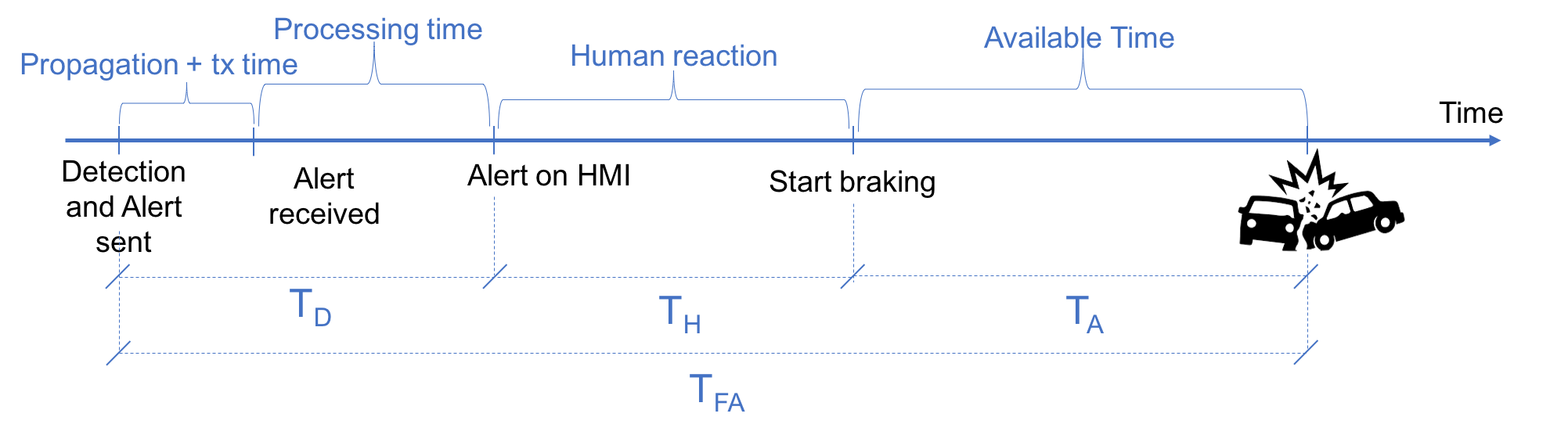}
\caption{Timeline of the communication between the detection server and the human driver.}\label{img:timeline}
\end{figure*}

\subsection{Reaction to Alerts} 
Whether a collision is detected in time or too late is
determined in the post-processing phase, by considering the alert messages  that have been received. A collision is considered as ``detected too late'' if:
$$T_A < T_B$$
where $T_A$ represents the time available to the driver to avert the
collision, i.e., the interval between when the driver initiates
evasive actions and the actual collision. $T_B$, instead, is the time
needed by the entity to stop, given its current speed and maximum
deceleration. $T_A$ is computed as follows: 
\begin{equation}\label{eq:TA}
T_A = T_{FA} - T_D - T_H \,,
\end{equation}
where the three elements in the above expression are:
\begin{itemize}
\item $T_D$: the time gap between the moment at which a collision is
  detected by the server and the moment at which the alert reaches the
  driver through the vehicle HMI. It includes the \textit{transmission
    time} and the \textit{processing
    time}. The transmission time includes the time to transfer data
  from the application server to the eNB, and then to the entities
  involved in the collision. The
  processing time is the time needed at the receiving node to process
  an alert message from the time instant at which the first bit is
  received to the time instant at which the driver is notified about
  the received information. The processing time is set to
  400\,ms \cite{5G-CAR};
\item $T_H$: the time needed by a human driver to take action
  following the prompt of an alert. It is fixed to 1 second, as suggested
  by several studies \cite{reaction1} \cite{reaction2} that take in
  account different variables such as age, travel length, environment,
  etc. Note that this parameter is set to zero in the case of
  automated vehicles; 
\item $T_{FA}$: the time interval between the generation of the first alert related to a possible collision and when the actual collision occurs.
\end{itemize}
Fig.~\ref{img:timeline} provides a visual representation of the
timeline of the communication between the collision detector and the human driver, highlighting the time intervals discussed above.

\section{Performance Results} \label{sec6}

Below, we describe the simulation settings we used (Subsection \ref{sub6.1}),
and we show our results in terms of collision detection
accuracy and alert reliability (Subsection \ref{sub6.2}). 

\subsection{Simulation Settings} \label{sub6.1}
We run two sets of ten 300s-long simulations, one with the server at the Metro node
and the other with the server placed in the Cloud. 
As exemplary values reflecting real-world mobile operators topologies, 5\,ms and 20\,ms have been chosen for the Metro-eNB and cloud-eNB latencies.
In the post-processing phase, each  of the two simulation sets is
analyzed considering either the human driver or the automated vehicle
case. Intuitive, a better performance can be expected in the automated
vehicle scenario, as $T_H=0$. 

\subsection{Simulation Results} \label{sub6.2}

\begin{figure}[t!]
\centering
\includegraphics[scale=0.39]{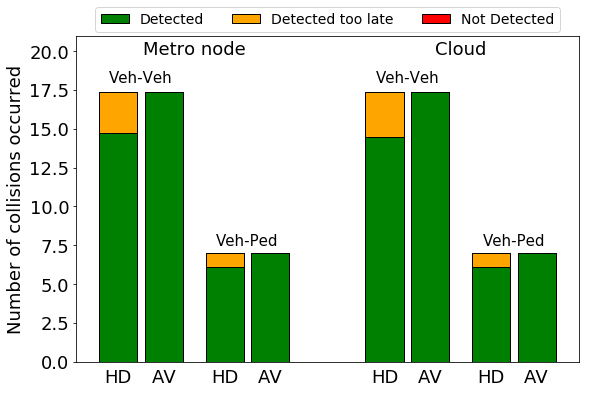}
\caption{Vehicle-with-vehicle (Veh-Veh) and vehicle-with-pedestrian (Veh-Ped) collisions detected, detected too late and not detected, for the Human Driver (HD) case and the Autonomous Vehicle (AV) case, and for different server placements.}\label{img:collisions}
\end{figure}
Fig. \ref{img:collisions} shows the effectiveness of our collision
avoidance system in terms of number of accidents that can be prevented. The four bars show the number of vehicle-with-vehicle detected, late-detected and undetected collisions among those reported by the SUMO simulator. A collision is reported in SUMO each time the polygon describing an entity overlaps with the polygon describing another entity. 
The two leftmost bars refer to the case in which the server is
placed at the Metro node, both in the human driver case and in the
automated vehicle case, while the other two refer to the scenario where
the server is in the Cloud.

\begin{figure}[t!]
\centering
\includegraphics[scale=0.4]{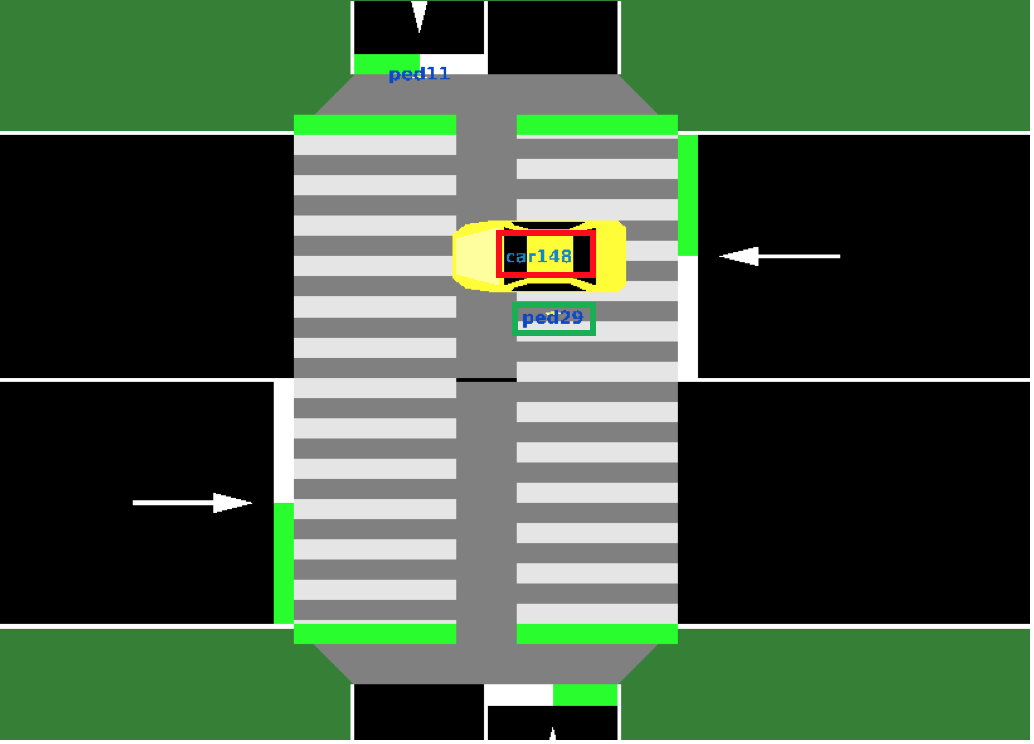}
\caption{Screenshot from SUMO representing a situation that leads to a false negative.}\label{img:pedOnCar}
\end{figure}

The first important result that we highlight is the effectiveness of our
algorithm: regardless of the location of the server, it reaches $100\%$
in case of automated vehicle, and over $80\%$ in case human driver. A
second relevant result, is the absence of ``Not detected'' collisions in
the four case studies. More in general, the histograms referring to the
Metro node are quite similar to the ones referring to the Cloud, with
the exception of few ``Detected too late'' cases: the latter show a
little increase when the server is placed in the Cloud. The reasons for this limited  increase are twofold. 
First, when a collision is correctly detected, the time
between the generation of the first alert and the actual collision
($T_{FA}$) is around 10 seconds (which is the threshold $t2c_t$). 
Thus, relatively speaking, the delay introduced by moving the
collision detector from the Metro node to the
Cloud can be considered as negligible. Second, even if the $t2c_t$
value were reduced, the increased transmission time would not have an
impact. Indeed, looking at the
expression of 
$T_A$ in (\ref{eq:TA}), the difference of 15\,ms due to the location of the
server, is very small compared to the processing time
(400\,ms) or, in case of human driver, to $T_H$ (1\,s). 

Next, we  focus on the
collisions between pedestrians and vehicles. The generation rate of
pedestrians is relatively low (on average 1 pedestrian every 10\,s), so the number
of collisions observed is lower than in the
previous case. We notice a decrease in the
effectiveness of our collision detector, with about 6.5\% of 
collisions going undetected. Furthermore, false negatives now affect all the case studies, regardless of the
type of driver and the location of the server. 
Since the ``undetected''
collisions are very few, we scrutinized each of them. It turned out
that all false negatives are due to the mobility model used in  SUMO
for pedestrians at zebra crossings and for the approaching vehicles.  
Let us consider a
vehicle and a pedestrian approaching a free zebra crossing. Since no
pedestrian occupies the crosswalk, the vehicle proceeds at maximum
speed. Once the pedestrian enters the zebra crossing, the car, which now
sees the obstacle, starts to decelerate. If it is impossible to stop in
time, the ``best'' thing to do would be not to stop. However,
according to the SUMO mobility model, the vehicle always tries to stop
when approaching an occupied 
junction,
thus the vehicle continues to decelerate. Our algorithm, which is
aware of the
speed, distance and acceleration of the car, predicts that the vehicle will never
hit the pedestrian and, also, that when it stops completely, the
pedestrian will be behind the car, thus no alerts are sent. However,
as 
shown in Fig. \ref{img:pedOnCar}, the vehicle stops
in the middle of the crosswalk and the pedestrian, completing the
crosswalk, rather than dodging it, physically walks over it. This is
reported as a collision in the SUMO logs. 
By discounting this issue, the system
performance becomes comparable to that of  vehicular collision case,
as depicted in Fig.~\ref{img:collisions}. Looking at the histogram,
moving the server to a farther location only minimally affects the
``Detected too late'' cases. This is  because, whenever a possible collision is
correctly detected, the pedestrian will have about 5\,s (which is the
$t2c_t$ threshold for the pedestrian case) to act, which is more than
enough since pedestrians can stop almost instantaneously.

\begin{figure}[t]
\centering
\includegraphics[scale=0.39]{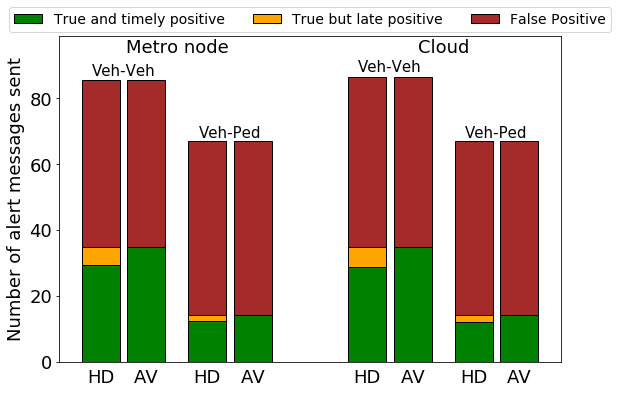}
\caption{False positive statistics for vehicle-with-vehicle (Veh-Veh) and vehicle-with-pedestrian (Veh-Ped) collisions
for the Human Driver (HD) case and the Autonomous Vehicle (AV) case, and for different server placements.}\label{img:alert}
\end{figure}
Another issue to investigate is the study of the {\em quality} of alerts that are actually received by the vehicles, in order to find the fraction of false positives, i.e., the alert messages referring to situations of low or no danger. False positives are not as critical as undetected collisions but they may be a cause of distraction for human drivers. The list of the metrics studied in this section follows:
\begin{itemize}
\item \textit{total alert sent}: total number of alerts sent by the server to vehicles to warn them about detected collisions.
\item \textit{true positives}: alerts that have been sent and refer to collisions that actually occurred. They include:
\begin{itemize}
\item \textit{true and timely positives}: alerts for which the driver had enough time to brake before the collision happened;
\item \textit{true but late positives}: alerts for which the driver did not have enough time to brake before the collision happened.
\end{itemize}
\item \textit{false positives}: alerts that have been sent and refer to collisions that would not take place.
\end{itemize}
Fig. \ref{img:alert} shows the results of this analysis. The false positive percentage is high, greater than 60\%. As was explained in Section \ref{sec4.3}, the value of false positives may decrease by decreasing the values of the thresholds $t2c_t$ and $s2c_t$. 

\begin{figure}[t!]
\centering
\includegraphics[scale=0.5]{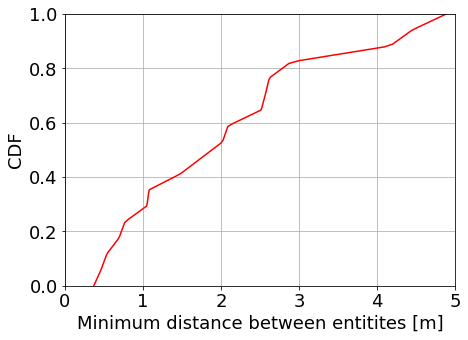}
\caption{CDF of the distance between cars in a false positive situation.}\label{img:cdf_car-car}
\end{figure}

This high value of false positive alerts brings us to analyze their
{\em relevance}. We therefore look at the minimum distance reached by
the pairs of vehicles receiving false positives alerts to determine
if, also in absence of collision, a dangerous situation has arisen. By
parsing the \textit{Floating Car Data} SUMO output, we checked, for each pair of vehicles warned by a false positive alert, the minimum distance that the two vehicles reached over the time. The Cumulative Density Function (CDF) of this quantity is shown in Fig.~\ref{img:cdf_car-car}.
This plot tells us that:
\begin{itemize}
\item 60\% of the false positive alerts are sent to vehicles that got less than 2.3\,m apart; 
\item there are no vehicles warned by a false positive alert whose minimum distance is more than 5\,m. 
\end{itemize}
This highlights that the false positive alerts are transmitted in
situations that are indeed dangerous -- a factor that is particularly
important when inaccuracies of the positioning system are taken into
account.

\begin{figure}[t!]
\centering
\includegraphics[scale=0.5]{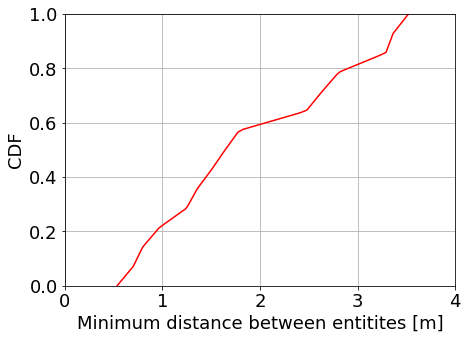}
\caption{CDF of the vehicle-to-pedestrian distance in a false positive situation.}\label{img:cdf_ped-car}
\end{figure}

As far as the vehicle-with-pedestrian false positive alerts are
concerned, the results are shown in Fig.~\ref{img:alert}. The
histograms portray a situation that is more critical than in
vehicle-to-vehicle collisions. Indeed, the false positive rate is around
80\%. This behavior is due to the characteristics of the pedestrian
mobility with respect to cars: zebra crossings are occupied for a longer
time by pedestrians, in particular, since pedestrians move
at a maximum speed of 2\,m$/$s and the two lanes are each 6\,m wide, a
pedestrian will occupy the crossing for about 6 seconds. During this
time, it is likely that other cars will approach the crossing and, if
they stop or pass close to the pedestrian, the collision detector will
trigger the generation of an alert, even if no collision actually
occurs. In this case too, it is important to look at the
CDF of the vehicle-to-pedestrian distance
(Fig.~\ref{img:cdf_ped-car}), which shows that 50\% of the false positives
are sent for situations in which the car and the pedestrian were 2\,m
apart from each other.

As a final remark, although the percentage of false positives is quite
high, a C-V2I-based system can ensure high reliability in collision
detection, and even false positives refer to actually dangerous
situations. Furthermore, large margins of improvement are possible if additional
information coming from on-board sensors (cameras, radars, lidars...) is
merged with that available through the C-V2I interface, and advanced
data fusion algorithms are used so as to provide the driver with a
comprehensive, yet accurate, warning system.

\section{Discussion and Conclusions} \label{sec7}
In this paper, we have proposed an efficient C-V2I-based system for
automotive collision avoidance, 
and tested it under different scenarios. 
By exploiting the transmission of CAMs toward the collision detection
server, the latter determines whether any pair
of  vehicles, or vehicle and pedestrian, are set on a
collision course, and, if so, it issues an alert message. We deployed the server in two different points of the
network, namely, in the Metro node and in the Cloud, and 
we considered both human drivers and automated vehicles.

Our results show that in every case study we analyzed, the percentage of
detected collisions is $100\%$. However, considering a human driver, a
percentage of those collisions (on average $14\%$) is detected too late. Considering  automated vehicle
instead, due to the absence of the human reaction time, no
late-detected collisions are observed. Additionally, in  this case, the different location
of the server does not have a noticeable impact on the system
performance, given the values for human reaction time and on-board processing time.

As far as the analysis of alert messages sent by the server is
concerned, we see that a high percentage thereof are false positives.
At a closer inspection, we found
that, even if the number of false positive alerts is very high, 
they are sent in situations that are actually dangerous. 
This fact is particularly important if inaccuracies of the positioning
system are taken into account.

Two main directions for future research can be envisioned: (i)
investigating the gain that cellular vehicle-to-vehicle (C-V2V)
communications can bring; (ii) assessing the benefits that may come
from data fusion performed on CAMs and sensory data such as those
collected via cameras, radars and lidars. 

\section*{Acknowledgment}
This work was partially supported by TIM through the research contract ``Multi-access Edge Computing'', and by
the European Commission through the H2020 5G-TRANSFORMER project (Project ID 761536).

\bibliographystyle{IEEEtran}
\bibliography{refs.bib}

\end{document}